# Progress with Xenon Liquid Hole Multipliers

Lior Arazi, Eran Erdal, Yevgeniya Korotinsky, Michael Rappaport, Arindam Roy, Sergei Shchemelinin, David Vartsky and Amos Breskin

*Abstract*– The bubble-assisted Liquid Hole Multiplier (LHM) is a recently-proposed concept for the combined detection of ionization electrons and primary scintillation photons in noble-liquid time projection chambers. The LHM comprises a perforated micro-pattern electrode (e.g. Thick Gas Electron Multiplier – THGEM, or Gas Electron Multiplier - GEM) immersed in the liquid, with a bubble of the noble gas supported underneath. Ionization electrons and scintillation-induced photoelectrons extracted from a cesium iodide photocathode drift through the electrode's holes and induce electroluminescence (EL) signals in the bubble; these are recorded by photon detectors located closely below the electrode. We present recent results in the development of LHMs, comparing the response of different electrodes to ionization and photon-induced electrons.

## I. Introduction

LIQUID Hole-Multipliers (LHMs) were recently proposed as a new detection concept of both VUV-photons and ionization electrons induced by particle interaction in noble liquids [1]. The original motivation was to find a solution to the challenge of maintaining a uniform electroluminescence (EL) response across the large diameter of future multi-ton dual-phase noble-liquid time projection chambers (TPCs) for dark matter detection; however, the concept may also be applicable in other fields, including neutrino physics, Compton imaging, and neutron detection. The original LHM concept consists of a perforated electrode, e.g., a Thick Gas Electron Multiplier (THGEM) or a Gas Electron Multiplier (GEM), immersed inside the noble liquid. Ionization electrons are focused by the field into the holes, where they induce EL signals. When coated with a cesium iodide (CsI) photocathode, the process can also occur for photoelectrons induced by VUV photons impinging on the LHM top surface. The EL signals can be recorded by small-pixel photodetectors, such as silicon photomultipliers (SiPMs) or gaseous photomultipliers (GPMs) located closely below the LHM.

Preliminary measurements with a THGEM electrode immersed in liquid xenon (LXe) [2] showed large EL yields at relatively low voltages. Later studies demonstrated that the process in fact occurs within a xenon *gas bubble* trapped under the electrode surface [3, 4]. This so-called bubble-assisted EL mechanism was found to be stable over months of operation, with up to 7.5% RMS energy resolution for ionization electrons from 5.5 MeV alpha particles stopped inside the liquid [4]. The estimated light yield in a THGEM was a few dozen photons per electron (emitted into 4π) at a THGEM voltage of 3000 V [3]. Additional studies extended the work to GEMs immersed in LXe, and provided first results of VUV photon detection using THGEM and GEM electrodes coated with CsI [5]. The present contribution includes further results on electron and VUV photon detection with CsI-coated THGEM and two types of GEM electrodes – a standard GEM (with bi-conical holes) and a single-mask conical GEM with larger hole-diameter and spacing.

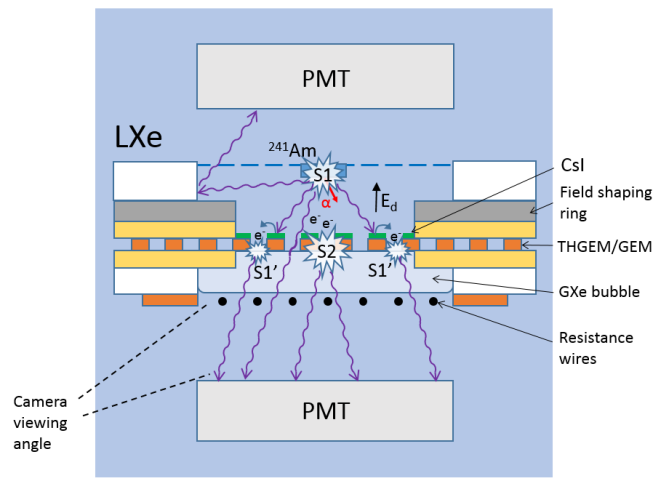

Fig. 1. Schematic view (not to scale) of the experimental setup for the recording of photoelectron and ionization-electron signals from a THGEM/GEM electrode immersed in LXe, comprising an $^{241}$Am alpha-particle source, field shaping ring, THGEM/GEM and resistance wire grating for generating bubbles. Signals are recorded from the bottom PMT, triggered on alpha-particle primary scintillation signals from the top PMT. An external camera views the THGEM/GEM and bubble from below. S1 is the primary scintillation signal, S2 is the EL signal induced by the ionization electrons and S1' is the EL signal induced by photoelectrons emitted from the CsI photocathode following the absorption of S1 photons.

## II. Experimental Setup

The study was performed using the LXe cryostat described in detail in [4]. The experimental setup is depicted schematically in Fig. 1. Its main components were an $^{241}$Am alpha particle source, a CsI-coated THGEM or GEM electrode, a grating of resistance wires for generating the bubble and two Hamamatsu R8520 PMTs at the top and bottom, operated at – 600 V in all measurements to be able to compare data from different experiments. An external CCD camera was used to

Manuscript received December 10, 2016. This work was partly supported by the Minerva Foundation with funding from the German Ministry for Education and Research (Grant No. 712025) and the Israel Science Foundation (Grant No. 791/15).

L. Arazi and E. Erdal contributed equally to this work. L. Arazi and M. Rappaport are with the Physics Core Facilities, Weizmann Institute of Science, Rehovot, 7610001, Israel (e-mail: lior.arazi@weizmann.ac.il). All other authors are with the Department of Particle Physics and Astrophysics, Weizmann Institute of Science, Rehovot, 7610001, Israel.

observe the electrode and bubble at an angle from below. The bubbles were either generated using the resistance heating wires or formed spontaneously by heat leaks as described in [4]. Either way, the EL signal magnitude and resolution were stable for days. In steady-state the lower bubble interface was at, or closely above, the plane of the wires. The bubble diameter, dictated by the spacer between the electrode and wire grating, was 30 mm, and its overall thickness ~2.5 mm.

A stainless steel ring between the source and THGEM/GEM electrode was used for field shaping, providing a nearly constant drift field along the axis of symmetry. High voltage was applied separately to the source, field shaping ring, THGEM/GEM top and bottom and optionally also to the wire grating.

Three different electrodes were used: (1) a THGEM with 0.3 mm diameter holes drilled in 0.4 mm thick FR4 at 0.7 mm pitch with 0.1 mm etched rims; (2) a standard GEM with bi-conical holes (top and bottom diameter 70 µm, central diameter 50 µm), etched in 50 µm-thick Kapton with a pitch of 140 µm; (3) a single-mask conical GEM ("single-conical GEM") with large holes (top diameter 150 µm, bottom diameter 190 µm), etched in 50 µm-thick Kapton with a pitch of 300 µm. In all cases the hole-pattern was hexagonal, with the Cu layers coated by Au. The central 14 mm diameter region of all three electrodes was coated with CsI (~300 nm thick) with a measured QE of ~20-22% in vacuum at 175 nm, matching in size the inner diameter of the field shaping ring.

The study aimed at comparing the relative EL light yield of the three electrodes in response to alpha-particle-induced ionization electrons and primary scintillation photons.

## III. RESULTS

We define three light signals (Fig. 1): S1 – the primary scintillation signal from the alpha particle track; S1' – the EL signal produced inside the electrode holes in response to photoelectron extraction from the CsI layer by S1 VUV photons; S2 – the EL signal produced inside the holes by the ionization electrons reaching the electrode from the alpha particle track. Fig.2 shows three typical waveforms including S1, S1' and S2 for the three different electrodes operated at or near their maximum stable voltage. Note that for the THGEM (A) and single-conical GEM (C) the bottom PMT was operated at −600 V to avoid saturation by large S2 signals, while for the standard GEM (B) the bottom PMT was operated at −800 V (with a ~10 fold larger gain). As can be seen in the Fig. 2, the largest S1' (relative to S2) is obtained for the single-conical GEM, followed by the standard GEM with the THGEM lagging behind. The interpretation for this is that for the two GEM electrodes the surface field is considerably larger than for the THGEM, providing a larger extraction efficiency of photoelectrons into the liquid. The improved S1' response (relative to S2) of the single-conical GEM, compared to the standard GEM, is likely the result of improved photoelectron collection into the holes.

Fig. 3 shows the dependence of S2 on the voltage across the three electrodes for a fixed drift field of 0.5 kV/cm. The single-conical GEM provides up to ~6-fold more light than the standard GEM for the same voltage, but similar light output as the THGEM at higher voltages. The differences in light yield between the three electrodes stem from differences in the local field, efficiency of electron collection into the holes, degree of bubble penetration into the holes, and solid angle for the EL photons emitted from within the holes towards the bottom PMT. Note that in all cases the S2 curve departs from linearity at large voltages, indicating the onset of avalanche gain inside the bubble (this is particularly evident for the single-conical GEM and THGEM).

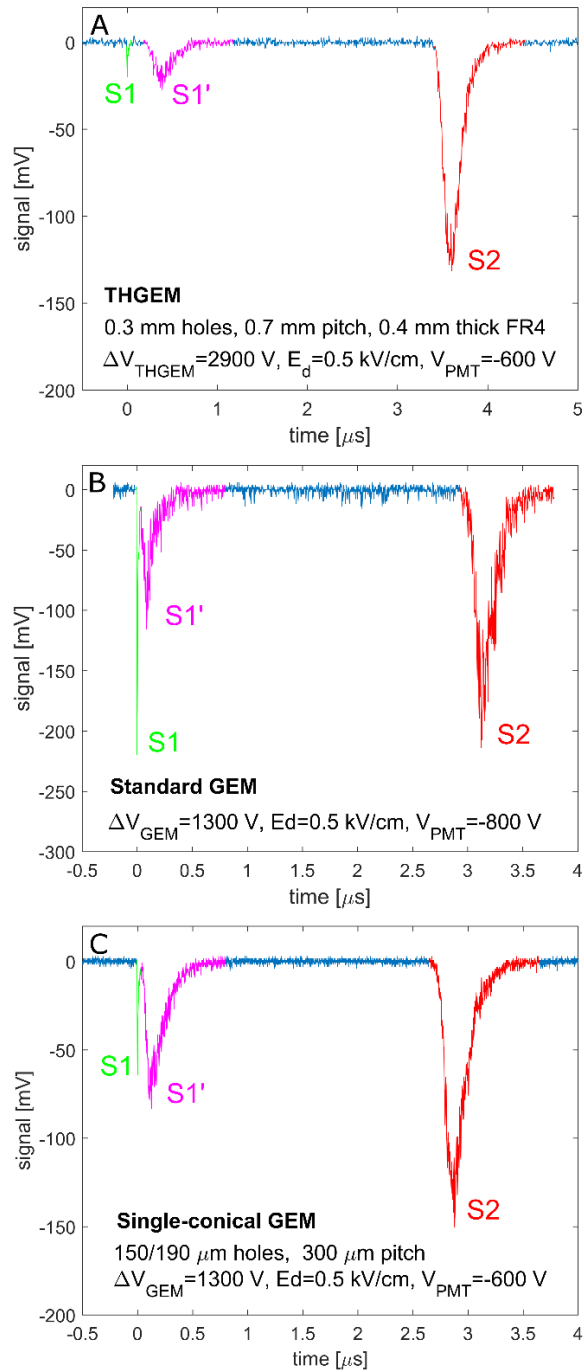

Fig. 2. Typical alpha-induced single-event signals for the THGEM (A), standard GEM (B) and single-conical GEM (C).

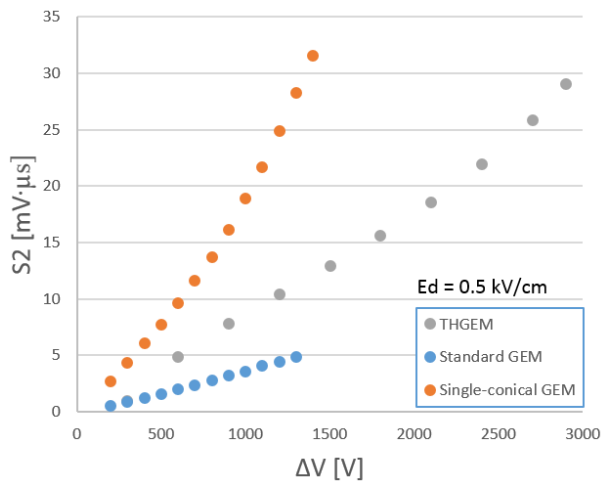

Fig. 3. S2 magnitude (pulse area) as a function of the voltage across the THGEM, standard GEM and single-conical GEM for a drift field of 0.5 kV/cm.

Fig. 4 displays the dependence of S1' on the GEM/THGEM voltage for a drift field of 0.5 kV/cm, showing a striking advantage of the single-conical GEM over the two other electrodes.

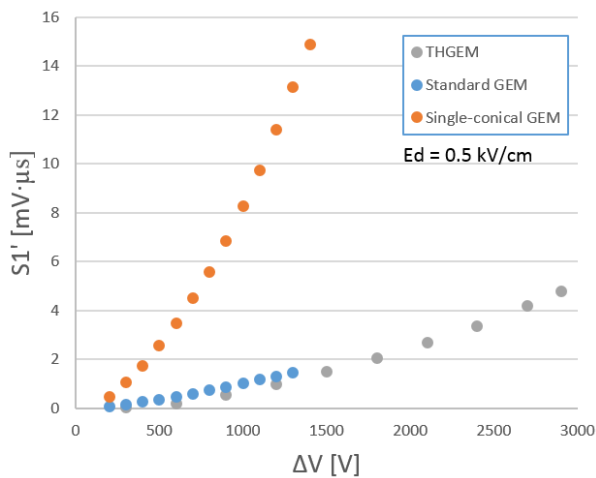

Fig. 4. S1' magnitude (pulse area) as a function of the voltage across the THGEM, standard GEM and single-conical GEM for a drift field of 0.5 kV/cm.

Fig. 5 shows the spectrum of the 5.5 MeV alpha particle S1' signal obtained by the single-conical GEM, operated at 1000 V with a drift field of 0.5 kV/cm. The estimated number of photoelectrons was ~4000. The RMS resolution, 5.1%, is the best value achieved so far with LHMs. The S1' RMS resolution obtained (for a drift field of 0.5 kV/cm) with a standard GEM was ~8% and with a THGEM ~6%. For a rough comparison, the RMS resolution of the EL signal in XENON100 for 4000 electrons was ~13% [6]. The S2 RMS resolution (at 0.5 kV/cm, with ~10,000 ionization electrons) was ~8% for all electrodes. The S1' resolution is considerably better than that of S2, in spite of the 2.5-fold smaller number of electrons. A plausible explanation is that S2 signals comprise a fluctuating contribution from gamma-rays and electrons emitted in coincidence with the alpha particle from the $^{241}$Am source; while their total energy is small compared to that of the alpha particle, their contribution to the number of ionization electrons is not negligible, because of the much larger fraction of electrons escaping recombination.

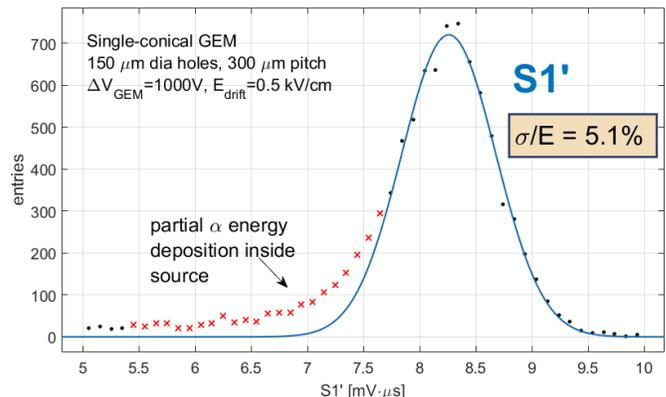

Fig. 5. S1' spectrum obtained using the single-conical GEM for 5.5 MeV alpha particles. The estimated number of photoelectrons is ~4000.

## IV. SUMMARY AND DISCUSSION

The present work focused on a comparative study of three different CsI-coated THGEM and GEM electrodes used as bubble-assisted LHMs in LXe, for the detection of both ionization electrons and primary VUV scintillation photons. Its main finding was that the single-conical GEM, with 150 μm top-diameter holes at 300 μm spacing, performs significantly better than both the standard GEM and THGEM tested in comparison. The single-conical GEM showed a several-fold larger EL yield for both ionization electrons and VUV scintillation-induced photoelectrons (at a given voltage), as well as better RMS resolution (in particular for the S1' signal). It is thought that the improved ratio between S1' and S2 for the single-conical GEM compared to the standard GEM stems from a better collection efficiency of photoelectrons into its holes. It is therefore expected that a much higher photon detection efficiency (PDE) can be achieved in this configuration. Further studies are needed to determine the optimal electrode geometry (larger holes may provide still higher electron collection efficiencies), as well as determine and optimize the PDE.

Parallel ongoing studies, to be discussed elsewhere, have recently demonstrated the feasibility of trapping a bubble behind vertical THGEM and GEM electrodes. This observation may open new possibilities for incorporating LHM modules in future noble-liquid detectors.

## ACKNOWLEDGMENT

The research was carried out within the DARWIN consortium for a future multi-ton LXe dark matter observatory. A. Breskin is the W.P. Reuther Professor of Research in The Peaceful Use of Atomic Energy.